\documentclass[aps,twocolumn,showpacs,nofootinbib]{revtex4}
\usepackage{graphicx}
\newcommand{\beq}{\begin{equation}}
\newcommand{\eeq}{\end{equation}}
\newcommand{\beqa}{\begin{eqnarray}}
\newcommand{\eeqa}{\end{eqnarray}}

\def\half{\frac{1}{2}}

\def\ket#1{|#1\rangle}

\def\opone{\leavevmode\hbox{\small1\normalsize\kern-.33em1}}

\begin{document}

\title{Are There Quantum Effects Coming from Outside Space-time?\\
Nonlocality, free will and "no many-worlds"}

\author{Nicolas Gisin \\
\it \small   Group of Applied Physics, University of Geneva, 1211 Geneva 4,    Switzerland}

\date{\small \today}

\begin{abstract}
Observing the violation of Bell's inequality tells us something about all possible future theories: they must all predict nonlocal correlations. Hence Nature is nonlocal. After an elementary introduction to nonlocality and a brief review of some recent experiments, I argue that Nature's nonlocality together with the existence of free will is incompatible with the many-worlds view of quantum physics.
\end{abstract}

\maketitle

\section{Introduction}\label{intro}
Imagine several persons that each separately and independently make choices that have consequences. For the sake of scientific analysis of this banal situation, assume that the same set of persons can repeat again and again the experiment, that is again and again make a free choice and observe its consequence. Moreover, for simplicity, assume that each one has a choice between a finite set of possibilities, that we name inputs, and that the consequences can be catalogued into a finite set of possible outcomes. Once enough data are collected, the probability of the various possible outcomes, given one possible input per person can be estimated. For example, if there are only two persons, that we may name Alice and Bob, and we label their inputs $x$ and $y$ and their outcomes $a$ and $b$, respectively, the probability reads: $p(a,b|x,y)$. For conciseness, we call such a conditional probability distribution, $p(a,b|x,y)$, a correlation, see Fig. 1.

Correlations are observed every day everywhere and, in particular, in all natural sciences. One could even argue that the scientific activity consists in observing correlations and developing theoretical models that explain them, i.e.~that describe how they happen. For example, if one watches a football game on a TV with the sound shut off and one observed that all the players simultaneously stop running, one would speculate, as an explanation based on our implicit theory of the game, that the umpire has whistled.

Surprisingly, the number of categories of explanations for correlations is extremely limited. Before quantum physics, there were only two categories of explanations: Either a first system influences a second one by sending him some information encoded in some physical systems, or the correlated events share some common causes in their common past. For example, in the football game, all players simultaneously stopped running because in their common past the umpire whistled, i.e.~acted as a common cause for all players.

The two categories of explanations are local in the sense that the processes start at a localized place and propagate locally from one place to an adjacent one. Hence, the usual terminology reads {\it local common cause}, to emphasize the central importance of locality which lies at the core of these explanations

It is difficult to imagine any other sort of explanation. Actually, if one insists that an explanation ought to be a kind of story that plays out in space and time, then I believe there is simply no alternative to the previously mentioned two categories of explanations. Yet, amazingly, quantum physics predicts entirely different kinds of correlations, called non-local correlations for reasons described below. Physics has a word for the cause of these non-local correlations: entanglement. But physics offers no story in space and time to explain or describe how these correlations happen. Hence, somehow, non-local correlations emerge from outside space-time (for an explanation of this provocative terminology see appendix A).

\begin{figure}[h]
\begin{center}
\includegraphics[width=1\linewidth]{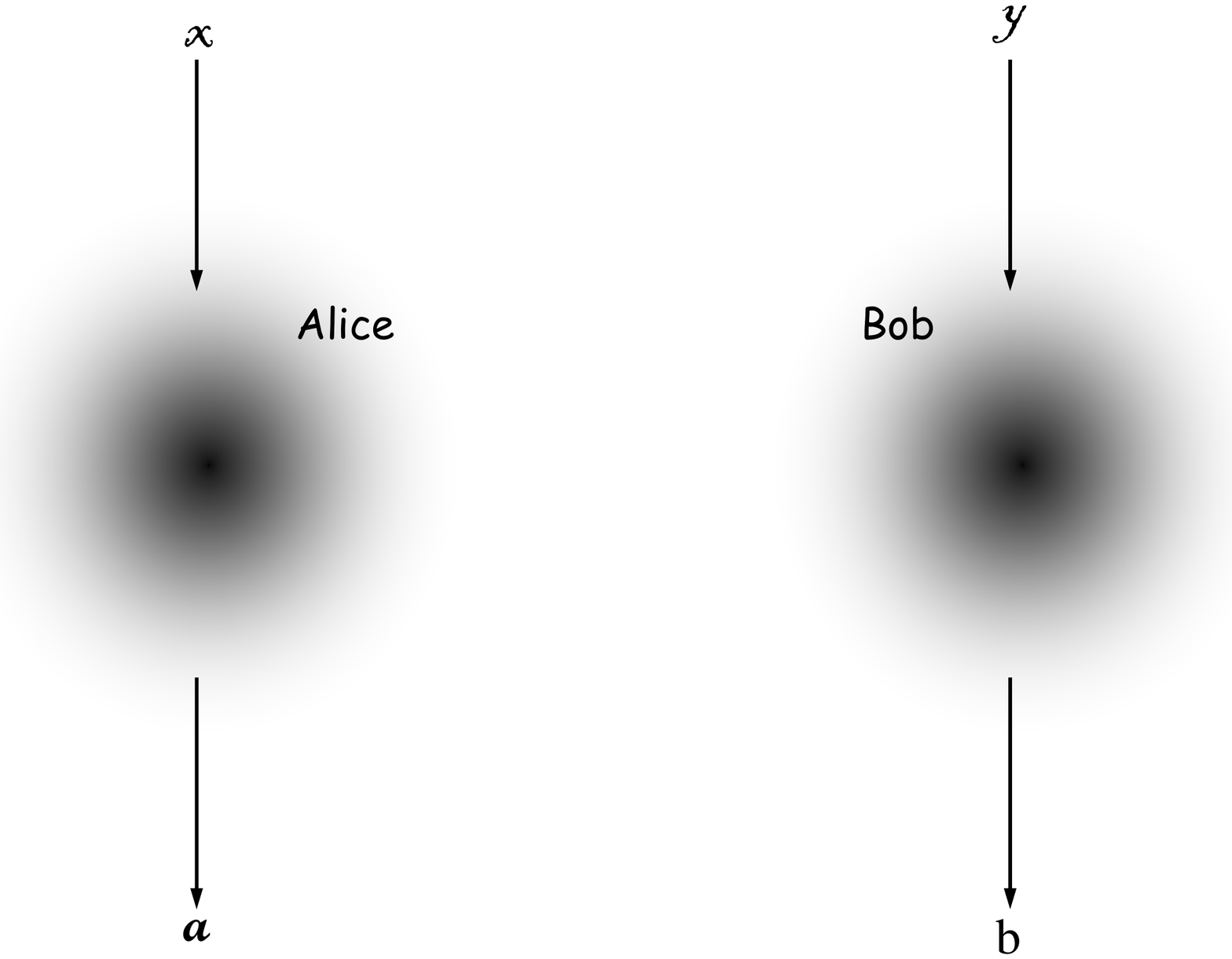}
\end{center}
\caption{\it For each run of the experiment, Alice and Bob each freely and independently chose one value $x$ and $y$, respectively, and input them into their black boxes; the latter then returns one and only one outcome $a$ and $b$ to Alice and Bob, respectively. Note that in order to test condition (\ref{CHSH1}), see section \ref{NLcorrel}, the experiment has to be repeated many times until the statistics allows one to infer a good approximation of the probability $p(a,b|x,y)$.}
\end{figure}

\section{Non-local Correlations}\label{NLcorrel}
Why should anyone believe the existence of nonlocal correlations? Their existence is predicted by quantum theory, as Einstein-Podolsky-Rosen and Schr\"odinger noticed already in 1935 (and actually years before, and also a few other precursors, see e.g.~\cite{Gilder}). But the possibility to directly observe nonlocal correlation seems, at first sight, difficult, if at all possible: indeed, one should observe correlations while simultaneously excluding any explanation of the two categories mentioned in the introduction.

The first type of explanation, i.e.~a first system sends information to the second, is quite easy to control, professors do that all the time during exams: they make sure the students can't communicate. In this way professors guarantee that if the exam's results are correlated it is not because one student copied the other, but because they prepared the exam together (i.e.~the only remaining explanation is local common cause). In physics, avoiding information exchange is straightforward, at least in principle: separate the correlated events such that nothing propagating at the speed of light can leave a system after the input has been given and reach the other before the outcome has been secured (one says then that the events are space-like separated). Let us emphasize this point. Bob should observe his outcome, i.e.~the consequence of his choice, before anything propagating at most at the speed of light could reach him carrying any information about Alice's choice; and vice-versa Alice should observe her outcome before any influence of Bob's choice, propagating at the speed of light, could reach her. Experiments that do not strictly fulfill this condition are said to suffer from the
"locality loophole".

But what about the second category of explanation, how could one experimentally rule out any local common cause explanation? The finding of a solution to this problem is John Bell's main contribution to physics \cite{Bell64}. It is pretty easy to formalize; let's have a look at Fig. 1. Alice and Bob should each have access to only a limited part of space and time. In particular one should be able to bound where and when the input choices are made (one by Alice, another one by Bob), and bound where and when the outcomes are produced and registered. Note that the inputs and outcomes are standard (i.e.~classical) variables: they can be copied, remembered and processed as any of the usual information we confront daily. For concreteness, assume Alice and Bob put their inputs and outcomes on the internet so that, after some time, everyone can access them. Let's assume that the correlation $p(a,b|x,y)$ has a common cause explanation. Let $\lambda$ denote this common cause. We do not need to know what $\lambda$ is, so far it is just a symbol. We make only two assumptions about $\lambda$, a serious one and a technical one. First the serious one: we assume that $\lambda$ doesn't contain any information about Alice's and Bob's free choice: the inputs $x$ and $y$ are independent of $\lambda$. Note that this excludes hyper-determinism: Alice and Bob can make truly free choices (I'll come back to this). This assumption can be formalized: $p(x)=p(x|\lambda)$. Or equivalently: $I(x:\lambda)=0$: the (Shannon) mutual information between $x$ and $\lambda$ is nil. The second assumption, the technical one, guarantees that one can "count" and "weight" all the possible common causes $\lambda$\footnote{Note that one can group the $\lambda$'s into equivalence classes where two $\lambda$'s are said equivalent iff they determine precisely the same probabilities $p(a,b|x,y,\lambda)$; hence it suffices that one can "count" the equivalence classes.}. A priori one doesn't know $\lambda$, but all that is necessary is to be able to associate probability weights to all the possible $\lambda$. For example, it suffices to assume that there are only countably many possible common causes, possibly infinitely countable (as the integers). Or, if one insists on the possibility of a continuous infinity of common causes (e.g.~the inputs depend on the temperature of some location in their common past), then one has to assume that the set of $\lambda$'s is equipped with a measure such that one can integrate over the space of $\lambda$'s \cite{Pitowsky}.

Now, if the correlation $p(a,b|x,y)$ has some local common cause explanation that satisfies the two above mentioned assumptions, then, for any given $\lambda$, the two events are independent:
\beq\label{indep1}
p(a,b|x,y,\lambda)=p(a|x,\lambda)\cdot p(b|y,\lambda)
\eeq
Since, a priori, one doesn't know $\lambda$ one has to attribute a certain probability to each of them: denote $\rho(\lambda)$ the probability that the actual common cause is $\lambda$. Note that the function $\rho(\lambda)$ may be unknown, but it is part of the local common cause category of explanations to assume that a $\rho$ exists. Consequently, any common cause explanation of correlations takes the form:
\beq\label{indep2}
p(a,b|x,y)=\sum_\lambda \rho(\lambda)~p(a|x,\lambda)\cdot p(b|y,\lambda)
\eeq
or if a continuous infinity of $\lambda$'s is assumed:
\beq\label{indep3}
p(a,b|x,y)=\int_\lambda \rho(\lambda)~p(a|x,\lambda)\cdot p(b|y,\lambda)~d\lambda
\eeq

Agreed? The rest of the argument is elementary mathematics. In brief, not all correlations $p(a,b|x,y)$ can be put in the form (\ref{indep2}) or (\ref{indep3}). Hence, if one observed a correlation that can't be written as (\ref{indep2}) or (\ref{indep3}), one has observed a correlation that can't be explained by local common causes. John Bell introduced a simple inequality, now generalized to entire families of so-called Bell inequalities, that are necessarily satisfied by all correlations of the form (\ref{indep2}) or (\ref{indep3}) \cite{Bell64}. We'll soon see an example: (\ref{CHSH0}) and (\ref{CHSH1}). Hence a violation of a Bell inequality is the signature of a correlation that can't be written as (\ref{indep2}) or (\ref{indep3}).

At this point it is worth emphasizing the interpretation of $\lambda$. Historically the $\lambda$ were thought of as local hidden variables by physicists whose hope was to restore some sort of local classical physics. A more modern view consists in viewing $\lambda$ as the physical state of the systems as described by any possible future theory. Hence, the violation of a Bell inequality tells us something not only about today's quantum physics, but tells us also something about any possible future theory compatible with today's experiments. That today's experiments tell us something important about any possible future theory is a rare and remarkable fact! Note furthermore how unrestricted $\lambda$ is: it could be the state of the entire Universe, except that $\lambda$ can't determine Alice and Bob's input choices $x$ and $y$. In this sense it is not $\lambda$ that is especially local, all that is assumed local is that Alice's system is not influenced by Bob's distant choice and vice-versa that Bob's system is not influenced by Alice's choice.

As a simple example of a correlation that can't be explained by common causes, consider the case where Alice and Bob have only to carry out a binary choice that we label $0$ and $1$, i.e.~$x,y\in\{0,1\}$, and their outcomes are also binary: $a,b\in\{0,1\}$. Note that this is the simplest possible case: with fewer inputs there would be no choice at all and with fewer outcomes the choices would have no consequences. The example goes as follows. Alice's outcome $a$ is random: $p(a|x)=\half$ for all inputs $x$; similarly Bob's outcome is random: $p(b|y)=\half$ for all inputs $y$. But the two outcomes are correlated: whenever it so happens that Alice and Bob both made the choice 1, i.e.~$x=y=1$, then their outcomes always differ: $p(a\neq b|x=y=1)=1$, and for all other combinations of input choices the outcomes are always equal. Since $x=y=1$ if and only if $x\cdot y=1$, this simple correlation can be captured with a simple relation:
\beqa
x\cdot y=0 \Rightarrow a=b \label{PR1}\\
x\cdot y=1 \Rightarrow a\neq b \label{PR2}
\eeqa
Note that this relation can be cast into a simple equation $a+b=x\cdot y$ (addition modulo 2), hence nonlocality shouldn't be hidden behind complex mathematics: the concepts are complex, not the maths. Let's analyze this correlation and look for a local common cause explanation. For this purpose we consider the following figure of merit:
\beqa \label{CHSH0}
S&=&p(a=b|x=0,y=0)+p(a=b|x=0,y=1)\nonumber\\
&+& p(a=b|x=1,y=0)+p(a\neq b|x=1,y=1)
\eeqa
Any local common cause $\lambda$ should, for all possible choices $x$ by Alice define an output $a$ (or define a probability for the outcome $a$), and similarly for Bob. For instance, one of the possible $\lambda$ is such that $a=b=0$ whatever the inputs. For such a $\lambda$ our figure of merit S takes the value 3: the first 3 terms in (\ref{CHSH0}) take value 1, but the last one is 0. It is not difficult to analyze all possible deterministic $\lambda$ (those $\lambda$'s that determine one and only one outcome on each side for any possible inputs), indeed there are only $2^2\cdot 2^2=16$ such $\lambda$'s. Analyzing these 16 $\lambda$'s one can easily convince oneself that our figure of merit $S$ never reaches a value larger than 3. And non-deterministic $\lambda$'s will not perform better (note that they can always be analyzed as statistical mixtures of the 16 deterministic $\lambda$'s). Consequently, all correlations explainable by local common causes satisfying the following inequality, named a Bell inequality:
\beq \label{CHSH1}
S\le 3
\eeq
Let me note for the more specialized readers that this inequality is strictly equivalent to the well known CHSH-Bell inequality: it suffices to note that the usual $E(x,y)\equiv p(a=b|x,y)-p(a\neq b|x,y)$ can equally be written as $E(x,y)=2p(a=b|x,y)-1=1-2p(a\neq b|x,y)$, the common form of the CHSH-Bell inequality follows then from (\ref{CHSH0}) and (\ref{CHSH1}): $E(0,0)+E(0,1)+E(1,0)-E(1,1)\le2$.

To conclude this section, let us emphasize the main conclusion: the two categories of local explanations for correlations can be experimentally tested. For this purpose one should observe correlations that violate some Bell inequality, as for example (\ref{CHSH1}), while making sure that the two observers, Alice and Bob, can't be influenced by any signal coming from the other side propagating at the speed of light (or slower). If such correlations are observed, there is no choice but to admit that there are correlations that can't be explained by any story in space and time. Such correlations are thus said to be nonlocal: there is no "local explanation", that is no explanation based on local causes that propagate from one place to adjacent ones.

\section{Experimental nonlocality}\label{expNL}
In this section I review some of the recent experiments, though without any of the important technicalities. Already in the famous Aspect experiment of 1982 the sides where space-like separated and the inputs chosen at "random" at the last moment so that no light-signal could explain the observed correlation \cite{Aspect82}. Admittedly there were no human Alice and Bob making free choices, only some pseudo-random, even somewhat periodic, choices where made by appropriate electronics. For scientists this was already extremely convincing, though since that time better experiments definitively closing the locality loophole have been performed \cite{LocLoopholeGeneva,LocLoopholeInnsbrug,ZbindenSwitch}.

All the above experiments observed correlations that violate a Bell inequality (\ref{CHSH1}). However, there is a little catch: in all experiments with photons (particles of light) there is often no outcome at all. For example, Alice inputs her choice, but nothing happens. Physicists understand why this is so, the photon got lost somewhere, or the detector supposed to register the tiny bit of energy carried by a single photon failed to do so (no real detector has 100\% efficiency), etc. Nevertheless, this is a serious loophole, called the detection loophole. Indeed, it could well be that the detection probability is influenced by the local common cause $\lambda$. Today, two experiments have closed this loophole using not photons, but ions \cite{DetLoopholeRowe,DetLoopholeMat}. This was a necessary step, however, in those two experiments the distance between Alice and Bob was insufficient to close the locality loophole. Hence, an experiment closing simultaneously the detection and the locality loophole is still awaited. Almost no physicist expects a surprise, certainly I do not expect any surprise, but the logical possibility remains and ought to be closed by further experiments.

So are we at the end? Do we have to conclude that Nature is nonlocal? Are there really correlations that can't be explained within space-time, i.e.~ that somehow emerge from outside space-time? The situation clearly deserves further scrutinies. In the remainder of this section I would like to analyze two local explanations together with experimental tests.

The first explanation is, I believe, very intuitive. Everything looks as if the two parties somehow communicate behind the scene \cite{BellHiddenComm}; hence, since they can't communicate at a speed equal or lower than the speed of light, let's assume they do so at a speed faster than light. Such an assumption doesn't respect the spirit of Einstein relativity, but this wouldn't be the first time that an accepted theory has to be revised \cite{incompleteRelativity}. Moreover, it is not crystal clear that such "communication behind the scene" would contradict relativity; indeed, one could imagine that this communication remains for ever hidden to humans, i.e.~that it could not be controlled by humans, only Nature exploits it to produce correlations that can't be explained by usual common causes. To define faster than light hidden communication requires a universal privileged reference frame in which this faster than light speed is defined. Again, such a universal privileged frame is not in the spirit of relativity, but also clearly not in contradiction: for example the reference frame in which the cosmic microwave background radiation is isotropic defines such a privileged frame.
Hence, a priori, a hidden communication explanation is not more surprising than nonlocality. It also has the very nice feature that it can be experimentally tested. The idea is to perform the measurements on both sides, i.e.~give the inputs and collect the outcomes, quasi-simultaneously. Hence, Bob's outcome can't be influenced by any hypothetical hidden communication and vice-versa for Alice's outcome. If the observed correlation is still nonlocal, i.e.~still violates Bell's inequality, then either the hypothesis of hidden communication is ruled out, or the speed of the hidden communication is faster than the bound set by the experimental condition, in particular by the accuracy of the synchronous timing and by the distance separating Alice and Bob. But there remains a conceptual difficulty: since we do not know which is the privileged reference frame, we do not know in which reference frame the event should be simultaneous. Philippe Eberhard suggested to exploit the rotation of Earth around its axes to scan all possible reference frame in 12 hours. This experiments has been carried out recently near Geneva \cite{SalartNature} and has set very stringent bounds on the speed of any hypothetical hidden communication: more than 10'000 or 100'000 times the speed of light, depending on technical details (see also the recent paper \cite{Cocciaro10}).

Before we come to the second alternative, let me mention that there is another way to define the faster than light hypothetical hidden communication: it could be that it is the inertial reference frame of the observer that determines that privileged frame. This very interesting idea was put forward by Antoine Suarez and Valerio Scarani in 1997 \cite{SuarezScarani97}. A consequence of this assumption is that, thanks to relativity, if the two observers Alice and Bob move apart fast enough, they could both, each in its own inertial reference frame, perform the measurement before the other, a so called before-before situation. This experiment was also be carried out in Geneva \cite{MovingObs}, and the observed correlation were still nonlocal: the proposal by Suarez and Scarani could be falsified.

The second way out of the conclusion "Nature is nonlocal" speculates on the fact that in actual experiments it is not so easy to determine when a choice is made and when an outcome is produced. Ideally, human Alice and Bob should make conscious choices, but in all experiments so far the choices are delegated to random number generators (or, even, no active choice is made, one merely argues - quite convincingly in my opinion - that the measurement settings are unknown to $\lambda$ to the particles until the moment they reach the measurement apparatuses). Delegating the choices to random number generator is pretty fine with me. After all, all what is required is that the choices are independent of the common past. Assuming that Alice and Bob's common past drives all choices made locally at Alice and Bob's locations by appropriate electronic or quantum devices seems to imply some sort of hyper-determinism that would make all Science an illusion (one could never decide to make an experiment, hence one could not test theories). Accordingly, let's concentrate on the idea that the outcome might, in fact, be determined much later than usually thought \cite{Franson85}. For example, two physicists, Lajos Diosi and Roger Penrose, independently proposed that an outcome is produced only once a mass has moved significantly (both proposed precise formulas relating the time of the outcome and the motion of the mass, their formulas agree within a factor 2 \cite{Adler07}). The motivation for this proposal lies in the difficulty to combine general relativity and quantum physics. But, never mind, here it suffice to note that in usual experiments the outcomes are collected in a computer's memories, hence without motion of any significant mass (electrons are very light). Hence, all observed violation of Bell inequality could be explained by slower than light influences: the influence has plenty of time to arrive before any mass moves significantly \cite{Kent05}. Fortunately, once again, this assumption of delayed outcomes can be experimentally tested. We coupled our detectors to a piezzo that could push a mirror and could thus falsify the Diosi-Penrose explanation of correlations violating Bell inequality \cite{Salart08}.

No doubt that further assumptions will appear. However, the huge amount of experimental data and the enormous predictive power of quantum physics very convincingly supports that Nature is nonlocal. So, how do physicist incorporate this amazing conclusion in their world view? Well, most simply don't care, most don't realize that they are living at a time of a huge conceptual revolution; sadly, most physicists would not have recognized Copernicus nor Galileo had they been contemporaries of these giants that carried out this conceptual revolutions. But there are exceptions that one may classify, roughly, in two categories: the many-worlds lovers and the others (to which I belong).

\section{Against Many-Worlds}\label{ManyWorld}
Basically the solution proposed by the many-worlds view of quantum physics, also called the multiverse, is to deny that experiments have unique outcomes \cite{KentManyWorlds}. According to this view, everything is quantum, once and for ever. Hence, the entire reasoning of section \ref{NLcorrel} collapses: there are no inputs and no outputs! Actually, the motivation for many-worlds is not nonlocality, but the fact that today's quantum theory offers no answer as to when a quantum measurement is finished. Hence, they conclude: quantum measurement are never finished, everything gets into an enormously complex state of superposition. Somehow, the only real thing is the Hilbert space and the linearity of Schr\"odinger's equation\footnote{Years ago, I once argued that the many-worlds doesn't seem compatible with Occam's razor principle \cite{GisinOccam}. As answer I got the following: "Occam's razor should not be applied to the physical world, but be applied to the Schr\"odinger equation; don't add any term to this beautiful equation" \cite{ZehOccam}. The linearity of the Schr\"odinger equation was assumed more real than our physical universe!}.

I won't try to present the many-worlds view any further; from the little above it should already be clear that I am not sympathetic with this view. But why am I so dismissive with this view while, at the same time, very open to all sort of assumptions like those presented in the previous section? Two reasons. First, all the assumptions presented in the previous section have an explanatory power. Moreover they could even be experimentally tested (and - even better for me - using technologies available in my lab!). On the contrary, I do not see any explanatory power in the many worlds: it seems to be made just to prevent one from asking (possibly provocative) questions. Moreover, it has built in it the impossibility of any test: all its predictions are identical to those of quantum theory. For me, it looks like a "cushion for laziness" ({\it un coussin de paresse} in French).

And there is a second, decisive, reason to reject the many-worlds view: it leaves no space for free will. I know that I enjoy free will much more than I know anything about physics. Hence, physics will never be able to convince me that free will is an illusion. Quite the contrary, any physical hypothesis incompatible with free will is falsified by the most profound experience I have about free will.

So, would I have rejected Newtonian classical mechanics had I lived before quantum physics? Probably not. Indeed, classical physics leaves open the possibility that free will can somehow interface with the deterministic Newtonian equations: free will could set-up some potential that could slightly influence particles's motion. This would be something like Descartes pineal gland. In standard quantum physics such an interface between free will and physics could be even simpler: free will could influence the probabilities of quantum events. This is, admittedly, a vague and not very original idea; but important is that there is no obvious definite contradiction between free will and standard quantum physics. However, the situation with the many-worlds view is very different.

In the many-worlds view all possibilities co-exist on equal footing. Accordingly, a being enjoying free will can't merely interact with one state of affair of the physical world, but has to interact with the enormous superposition of all possible states of affair! But, most likely, if a specific interaction with one possible state of affair produce a desired effect, this very same specific interaction with most of the other - equally real according to many-worlds - state of affairs would produce uncontrolled random effects. Hence, it seems that there is no way to maintain a possible window for free will in the many-worlds view. Consequently, I believe the many-worlds view is excluded by our daily experience.

A possible way out of the above reasoning could be to envisage that the being enjoying free will is also in an enormous superposition state and that the branches of this superposition match the branches of the superposition of the physical world. Hence, in each branch a story similar to the one sketched above in the case of Newtonian classical mechanics could hold (to maintain hope). But this "way out" is an illusion. Indeed, it would imply that the being enjoying free will actually never makes one and only one decision, nor experience one and only one consequence of his choice: he would make a superposition of all choices and experience all possibly consequences. In brief, such a being would enjoy no free will at all.

In summary, superpositions and entanglement forever, i.e.~the many-worlds "solution" to nonlocality, is not compatible with our most intimate experience as beings who enjoy free will. I make choices that have consequences; hence superpositions and entanglement must end somewhere. And the fact that today's physics doesn't know where they stop doesn't affect this conclusion at all.

\section{Conclusion}\label{Concl}
We have seen that any proper violation of a Bell inequality implies that all possible future theories have to predict nonlocal correlations. In this sense it is Nature herself that is nonlocal (section \ref{NLcorrel}). But how can that be? How does Nature perform the trick \cite{GisinScience}? Leaving aside some technical loopholes, like a combination of detection and locality loopholes, the obvious answer, already suggested by John Bell \cite{BellHiddenComm}, is that there is some hidden communication going on behind the scene. A first meaning of "behind the scene" could be "beyond today's physics", in particular beyond the speed limit set by relativity. We have seen how this interesting idea can be experimentally tested (section \ref{expNL}) and how difficult it is to combine this idea with no-signaling (appendix \ref{HcommLHV}). Hence, it is time to take seriously the idea that Nature is able to produce nonlocal correlations \cite{nonrealism}. There are several ways of formulating this:
\begin{enumerate}
\item Somehow God plays dice with nonlocal die: a random event can manifest itself at several locations.
\item Nonlocal correlations merely happen, somehow from outside space-time, in the sense that no story in space-time can describe how they happen (see appendix A).
\item The communication behind the scene happens outside space-time
\item Reality happens in configuration space, what we observe is only its shadow in 3-dimensional space (this might be the closest to the description provided by standard quantum physics) \cite{GisinBellPrize}.
\end{enumerate}

Admittedly, the situation is serious, so much so that despite the vast evidence further scrutinies should be undertaken. However, at this point we should have the courage to also seriously consider the possibility that Nature is indeed truly and deeply nonlocal\footnote{Many physicist hate this conclusion because they fear that it allow faster than light signaling. Hence, let me emphasize that nonlocality does not necessarily imply faster than light signaling. Actually, today's paradigm for most specialists is {\it nonlocality without signaling}}.

At this point one should ask oneself whether this is really new or whether similar conclusions already follow from the non-deterministic characteristic of quantum physics? Indeed, one could argue that non-determinism implies that the cause originates from elsewhere, i.e.~somewhere outside space-time. But this doesn't sound very convincing. I have no problem with the idea that certain objects may have an intrinsic propensity to spontaneously act in a stochastic manner. Furthermore, stochasticity by itself could act purely locally. Hence, with nonlocality we face something deeply different.

One logical possibility to avoid the entire argument - and hence the conclusion "Nature is nonlocal" - is to deny the possibility to freely choose inputs and/or collect measurement outcomes. One could invoke some hyperdeterminism such that the state of the universe $\lambda$ necessarily determines the inputs $x$ and $y$, but this seems to me like giving up the entire scientific enterprize. Indeed, with such a totalitarian determinism there would be no way to test one's scientific theories. Alternatively on could deny that measurements have outcomes, or at least that it takes in fact much longer for an outcome to be definitive than usually thought. An example, discussed in section \ref{expNL}, could be that a measurement outcome is definitive only once a mass has significantly moved. This interesting explanation of the observed correlation could be experimentally falsified. Another example could be that a measurement is finished only once a human becomes conscious of its outcome ... but then, as John Bell put it, "does that human need to have a PhD?". Clearly such ideas are ill defined, though they deserve further scrutinies. Finally, pushed to the extreme, one could argue with the many-worlds "lovers" that measurements don't have outcomes, that all possible outcomes remain potential in some huge superposition state containing all possibilities on an equal footing. I have argued in section \ref{ManyWorld} that such an extreme view is uninteresting and necessarily false because it is incompatible with free will. Admittedly, no physical theory so far has ever been able to include free will in an interesting way; however, the many-worlds view seems to be the first one totally incompatible with our most intimate experience of free will.

\appendix
\section{What could it mean that nonlocal correlations emerge from outside space-time?}\label{StoryToolBob}
In physics we develop mathematical models that allow us to compute the outcome of some experiments, or their outcome probabilities, i.e. we have equations. However, this is only half of theoretical physics. We also develop stories that describe how things happen. For example the moon attracts the water in the ocean, hence producing the tides. Or, we describe the relation between temperature and pressure of a gas by a story like: the gas is made out of trillions of little particles that move in all directions; the warmer the gas the faster the particles on average; when the particles hit the recipient containing the gas they exercise a small force on it, hence the trillions of particles all together exercise some pressure on the container; finally the pressure is larger when the average velocity of the particles is larger. Who has ever started a physics course with equation and not with a story? Clearly, in physics we need stories as much as equations. For this purpose we have a catalogue of possible tools to tell our stories. Until recently, all stories took place in space-time. But, this story-toolbox evolves as our theories evolve in parallel with our mathematics-toolbox; see for example the tools used today to talk about the deformation of space-time in general relativity.

However, as we have seen in scetion \ref{NLcorrel} no story in space-time can describe nonlocal correlations: we have no tool in our story-toolbox to talk about nonlocal correlations. Hence, we usually say things like "event A influences event B", or "event A has a spooky action at a distance on event B" or "event A causes a collapse of the wavefunction at location B". But we know that this is all wrong: there is no time ordering between the events A and B; hence no story in time is appropriate. Moreover, the distance between A and B is irrelevant; hence the distance should not occur in our story. The usual reaction to this situation is to give up the search for any story, i.e. in some sense to give up the very possibility to make sense of nonlocal correlations, i.e. to understand them. Some physicists simply claim that the maths are too complicated, hence we can't complement the equations by good stories. But we have seen that the maths are trivial: this can't be an excuse to give up!

\begin{figure}[h]
\begin{center}
\includegraphics[width=1\linewidth]{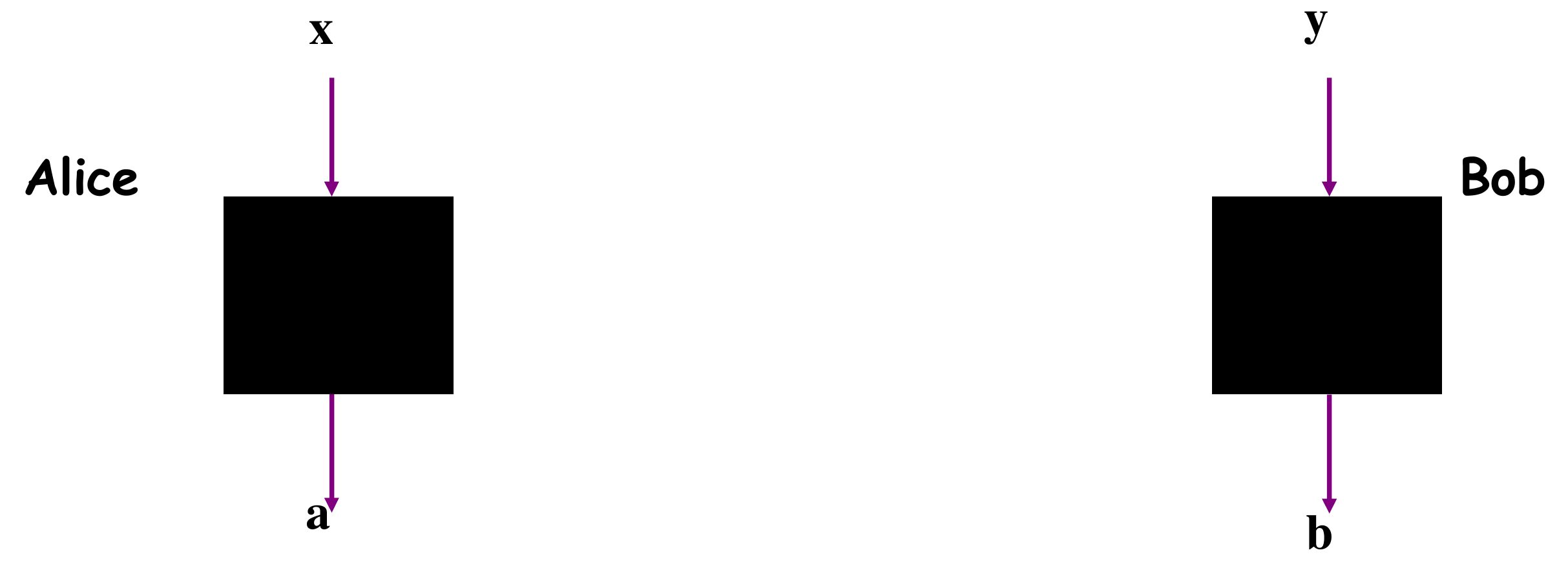}
\end{center}
\caption{\it Example of a possible new tool to talk about nonlocal correlations. The inputs $x$ and $y$ and the outcomes $a$ and $b$ are all bits. As soon as an input is fed into a box, the box produces a random outcome. But the outcomes hold the following promise: $a+b=xy$ (addition modulo 2). This promise holds irrespective to the inputs time ordering and independently of the distance between Alice and Bob. As explained in the text, this is a simple and powerful example of a nonlocal correlation.}
\end{figure}

Admittedly we need to enlarge our story-toolbox. A difficulty is that the new tool must include some strange features that can't be described within space-time. I am confident that with future quantum technologies this new piece in our story-toolbox will be familiar to future generations. Let me give an example of how this new piece could look. Imagine a pair of boxes connected by an immaterial link, see Fig. 2. Each box can be fed by an input, denoted $x$ for the first box and $y$ for the second; as soon as a box receives an input, it produces an outcome denoted $a$ and $b$ for the first and second box, respectively. For simplicity imagine that the inputs and outcomes are binary: both inputs and both outcomes are simply bits, i.e. a "0" or a "1". Locally the outcomes are random: locally the boxes just produce noise, but assume that the outcomes hold the following promises: $a+b=xy$ (addition modulo 2). This is identical to the relations (\ref{PR1}) and (\ref{PR2}). This new tool is unfamiliar to us, but it is quite simple. Moreover, it contains the essence: the promise that $a+b=xy$ holds irrespective of the timing where the inputs are given and holds independently of the distance between the two boxes; furthermore the correlation $p(a,b|x,y)$ is nonlocal (i.e. it can't be described by local common causes because it violates the Bell inequality (\ref{CHSH1})). This tool is well known to specialists and is referred to as a "nonlocal box" \cite{PRbox}. It shares, with quantum nonlocal correlations, the important feature that it can't be cloned \cite{PRnoCloning} (and the proof is very simple: again a nice story), accordingly one can also tell a simple story based on it about quantum cryptography \cite{Bell2QKD}. Finally, let me mention that with such a nonlocal box all quantum correlations corresponding to two maximally entangled qubits can be reproduced \cite{PRsinglet}, hence the nonlocal box contains enough nonlocality to encompass the most usual correlations one encounters in quantum physics. However, to be fair, I should add that this new tool is insufficient to tell a story describing quantum teleportation \cite{PRnoTeleportation}.

Hence, more tools are needed. Looking for such new tools, however, is not standard research in physics. Nonetheless, can we really expect physics to make progress and be appreciated by the public, as it should, if we give up the possibility to tell stories about it?

\section{Hidden Communication without hidden variables}\label{HcommLHV}
Experiments can only set bounds on the speed of any possible faster than light hidden communication. What about infinite speed? and could a theoretical argument refute the possibility of hidden communication at an arbitrarily fast but finite speed?

Let me first briefly comment on the idea of hidden communication at infinite speed. Frankly, I have difficulties making sense of such an assumption: essentially it implies that everything could instantaneously influence everything else \cite{Garisto}.

\begin{figure}[h]
\begin{center}
\includegraphics[width=1\linewidth]{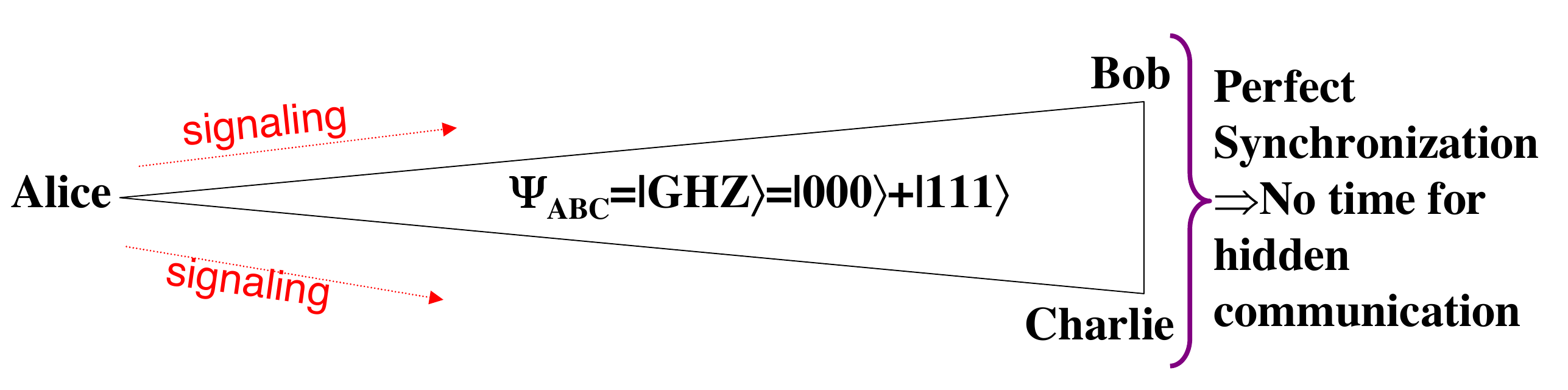}
\end{center}
\caption{\it In such a configuration, if all correlations are due to hidden communication behind the scene, then Alice can signal faster than light to Bob/Charlie.}
\end{figure}

Interestingly, however, the second question has at least a partial and positive answer. Indeed, one can prove that there are 3-party scenarios in which any explanation of distant correlations based purely on hidden communication (at any finite speed), hence without any additional local variable $\lambda$, would allow one to signal faster than light \cite{ScaraniGisinBrazil}. The argument runs as follows \cite{GisinBellPrize}. Imagine that the 3 players, Alice, Bob and Charlie, share a GHZ state of 3 qubits: $(\ket{000}+\ket{111})/\sqrt{2}$. Alice is far both from Bob and from Charlie. Bob and Charlie are not as far from each other, but still far enough that their input-outcome events are space-like separated, see Fig. 3. Further, imagine that Bob and Charlie synchronize their events so well that there is no time for the hidden communication to influence each other. Consequently, if Alice does nothing, but Bob and Charlie measure their qubits in the standard $\{\ket{0}$,$\ket{1}\}$ basis, then they observe random and uncorrelated outcomes. Indeed, all qubits are locally in a random state and there is, by assumption, no time for any influence (even at a speed possibly faster than light, but finite). If, however, Alice makes a measurement, also in the standard basis, long enough before Bob and Charlie (in the privileged reference frame) so that the hidden communication from Alice to Bob and to Charlie has time to arrive, then Bob and Charlie's outcome are correlated: they are both equal to Alice's outcome. Hence, if Bob and Charlie compare their results, they know whether Alice made a measurement or not, i.e.~there is signaling from Alice to (Bob,Charlie). Note that comparing Bob and Charlie's result takes some time, but since Alice could be arbitrarily far away, there is clearly a possibility that the signaling from Alice to (Bob,Charlie) is faster than light.

The above argument illustrates how difficult it is to modify quantum physics while maintaining nonlocality without signaling. However, the sketched argument is clearly of limited scope: it is easy to avoid signaling by adding some local variables $\lambda$ and by assuming that if the hidden communication doesn't arrive on time, then the outcomes are determined by these additional $\lambda$'s. It is thus desirable to extend the argument to include mixed models, that is a mix of hidden communication and additional local variables $\lambda$. It would be nice to show that any such mixed model necessarily activates signaling in some multi-partite scenarios. I find this research program highly interesting.



\end{document}